# Emergence of the interplay between hierarchy and contact splitting in biological adhesion highlighted through a hierarchical shear lag model

Lucas Brely, [a] Federico Bosia [a] and Nicola M. Pugno *[b, c, d]

**Abstract**

Contact unit size reduction is a widely studied mechanism as a means to improve adhesion in natural fibrillar systems, such as those observed in beetles or geckos. However, these animals also display complex structural features in the way the contact is subdivided in a hierarchical manner. Here, we study the influence of hierarchical fibrillar architectures on the load distribution over the contact elements of the adhesive system, and the corresponding delamination behaviour. We present an analytical model to derive the load distribution in a fibrillar system loaded in shear, including hierarchical splitting of contacts, i.e. a "hierarchical shear-lag" model that generalizes the well-known shear-lag model used in mechanics. The influence on the detachment process is investigated introducing a numerical procedure that allows the derivation of the maximum delamination force as a function of the considered geometry, including statistical variability of local adhesive energy. Our study suggests that contact splitting generates improved adhesion only in the ideal case of infinitely compliant contacts. In real cases, to produce efficient adhesive performance, contact splitting needs to be coupled with hierarchical architectures to counterbalance high load concentrations resulting from contact unit size reduction, generating multiple delamination fronts and helping to avoid detrimental non-uniform load distributions. We show that these results can be summarized in a generalized adhesion scaling scheme for hierarchical structures, proving the beneficial effect of multiple hierarchical levels. The model can thus be used to predict the adhesive performance of hierarchical adhesive structures, as well as the mechanical behaviour of composite materials with hierarchical reinforcements.

**Introduction**

Animal contact elements exploiting dry adhesion, such as those found in insects [1-2], spiders [3-4] or geckos [5-6] share a common strategy to enable optimized attachment to a non-adhesive substrate: contact is achieved through a large number of fibrillar structures that interact with the surface through van der Waals interactions [7] and/or capillary forces [8]. A large variety of behaviours have been observed [9], but in general the adhesive strength of the contact pads has been found to increase as the size of the terminal elements (i.e. spatulae or setae) decreases and their number increases [1]. Indeed, contact models such as that by Johnson, Kendall and Roberts (JKR) [10] predict an unlimited increase in the adhesive strength as the size of the contact tips decreases. This decrease in size also leads to an increase of the total peeling line, i.e. the sum of all contact tip widths, which is proportional to the peeling force according to thin-film peeling theories [11]. Additionally, as the size of the animal increases and the dimensions of the contact units are reduced, hierarchical splitting is observed. For example in geckos, the lamellae support so-called setae, which are themselves split into hundreds of spatulae [6]. Similar structures are observed in arachnids [4]. Fibrillar contacts have been shown to be beneficial over non-fibrillar ones in certain ranges of the mechanical parameters [12]. Additionally, the hierarchical arrangement of fibrillar adhesives has been described as a way to increase the work of adhesion [13], optimize surface adaptability [14] or self-cleaning abilities [15] and to avoid self-bunching [13], and has been extended not only to the hairy adhesive structures, but also to spider silk anchorages [16-18]. Frictional properties of adhesive systems have also been recently discussed [5, 19-20]. Despite these numerous works, important aspects remain to be discussed relative to the biological or artificial fibrillar adhesives, such as the influence of hierarchical structure on the load distributions to which the contact elements are subjected, or on the energy dissipation occurring during delamination. With the recent introduction of artificial micro-patterned surfaces that mimic animal adhesion [21-22], including hierarchical structures [23-24], reliable analytical and numerical approaches need to be developed in order to derive optimization criteria for such systems [25] or dependence on various parameters [26] and the interplay between contact size and hierarchical organization needs to be adequately addressed.

In this work, we present an extension of a classical shear-lag model to hierarchical configurations and introduce a numerical approach to simulate the detachment process of thin films with an arbitrary hierarchical structure from rigid substrates, with the objective of calculating the load distributions acting on their contact units, validating the theory and providing predictions for the peeling force of hierarchical adhesives

**Model**

**Thin film peeling**

*Figure.1.A* schematically illustrates a thin film, or tape, adhering to a substrate and the longitudinal and shear stress distributions $\sigma_l(x)$ and $\tau(x)$ occurring at the interface along an infinitesimal length *dx* when a load is applied in the vicinity of the detachment front, referred to as the "peeling line" [11]. The interface region where these distributions occur is referred to as the "process zone" [27]. Kaelble proposed to model the film deformation by assigning it a finite axial, bending and shear stiffness, in order to study the problem in terms of an elastic beam on an elastic foundation [28]. He proposed to use a differential beam and adhesive element to extract these distributions analytically, relating them to strain energy release considerations. Considering that the detachment propagation of an adhesive tape is a mixed mode fracture problem involving normal (mode I) and tangential (mode II) load to failure, the peeling front propagates when:

$$G_I + G_{II} > G \tag{1}$$

where $G_I$ and $G_{II}$ are the strain energy release rates corresponding to mode I and mode II failure, and $G$ the adhesive energy available at the interface between the tape and the substrate. Kendall also used energy balance criteria to analytically describe the delamination ("peeling") of a tape from a substrate, and developed a general model for $G$ in the case of a thin-film geometry [29]. In his model, detachment occurs when

$$G = \frac{F_C}{w}(1 - \cos\theta) + \frac{F_C^2}{2Ebw^2} \tag{2}$$

where $F_C$ is the detachment force, $w$ the tape width, $b$ the tape thickness, $E$ the tape elastic modulus and $\theta$ the angle between the load direction and the substrate, referred to as the "peeling angle". Only when the load is parallel to the substrate, the adhesive energy coincides with mode II strain energy release rate, i.e. $G_I = 0$ and $G = G_{II}$, so that only the tangential forces along the interface are responsible for the adhesive interface failure, with:

$$G = \frac{F_C^2}{2Ebw^2} \tag{3}$$

In this case, the strain energy release rate is only linked to the recoverable work of the deformable tape under tension [27]. For stiff tapes, (i.e. $E \to \infty$), as the peeling angle increases, the normal distribution becomes more critical and for large $\theta$ values, the strain energy release rate is mostly influenced by the non-recoverable work due to the advancing peeling line:

$$F_c \cong \frac{wG}{(1 - \cos\theta)} \quad (4)$$

The latter equation is usually associated with the Rivlin model [30], which provides the peeling force of an inextensible tape as a function of the adhesive energy.

Here, we consider the case where the tangential forces at the interface are mainly responsible for the detachment, i.e. we focus our analysis for small peeling angles. As shown in *Figure.1.B*, in this case the strain energy release rate of the problem tends to Eq. (3). In this case, only the axial load of the attached tape structure transferred trough the interface layer is considered and the force balance can be reduced to a 1-D problem, usually referred to as the "shear-lag model" [31], leading to a simple description of the load distribution. This loading configuration corresponds to the case in which the detachment force reaches its maximum, and is representative of the loading condition acting on biological contact elements (e.g. a gecko toe pad) in a stable attached configuration. Indeed, it has been shown that animal attachment systems [27, 32-34] take advantage of the increased adhesive strength at small peeling angles. Geckos, for example, use opposing legs to stick to a surface in an inverted, upside-down position, thus reducing the peeling angle and optimizing adhesion.

Kaelble [28] extracted the exact shear distribution from the tape/interface shear lag model, which also allows to obtain the mode II strain energy release rate from the finite shear stress level at the peeling line:

$$G_{II} = \frac{F_c^2}{2Ebw^2} \cos^2\theta \quad (5)$$

From (Eq.2) and (Eq.5), the mode I strain energy release rate can be obtained:

$$G_I = \frac{F_c^2}{2Ebw^2} \sin^2\theta + \frac{F_c}{w}(1 - \cos\theta) \quad (6)$$

Figure 1.C shows the contributions of the two failure considered modes in the peeling test, showing that shear failure is dominant at small angles, and in general within the entire range of peeling angles observed in animal adhesion ($\theta < 20°$).

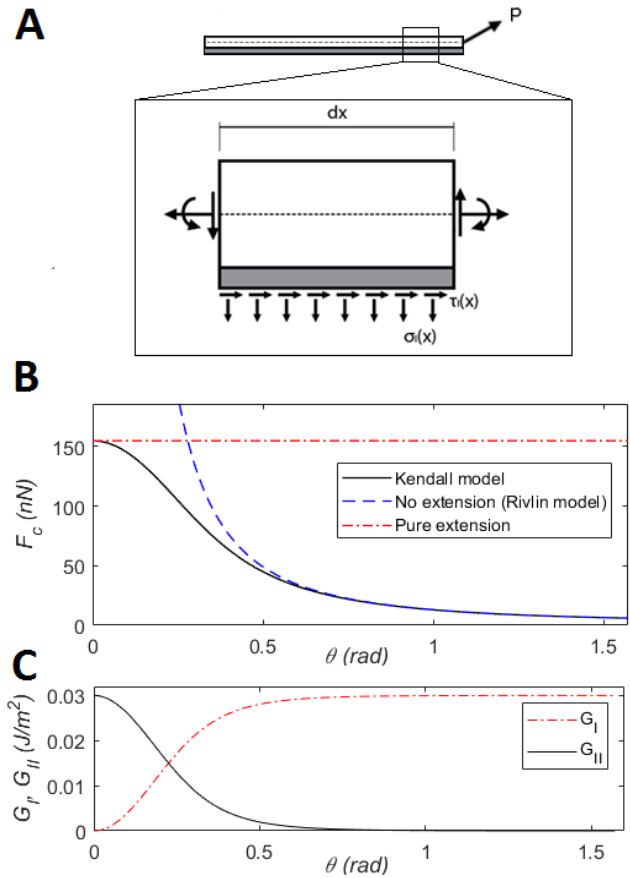

*Figure 1 : A. differential beam element used in [24] to extract normal and shear load distributions at the interface between the tape and the substrate. B. Peeling force vs. angle for various models: Kendall's model (Eq.(2) ), Rivlin model (Eq. (4)) and the peeling force limit in pure extension ($\theta = 0$, Eq. (3)). C. Mode I and mode II strain energy release rate at detachment as a function of the peeling angle.*

**Hierarchical Shear-Lag Model (HSLM)**

A schematic of the considered hierarchical attachment system geometry is given in *Figure.2.A*. For the reasons explained above, we now focus our study to the case of a load directed parallel to the substrate, since this provides significant insight in the role of hierarchy and contact splitting, starting from the analysis of the corresponding load distributions, and their influence on delamination. Rather than directly transferring the load between the tape (level-*h* structure) and the interface, intermediate structures are introduced (level-(*h*-1) , … , level-1, level-0) in the form of arrays of smaller tapes. The stress is transferred to the substrate only through tape-like contacts that support axial stress only, according to a Kendall model description. The attachment system thus becomes a self-similar structure that transfers load through hierarchically organized contact units. The force acting on an infinitesimal length *dx* of the level-*h* tape is shown in.*B*. At each scale level-*h*, the tape geometrical and mechanical properties are the width $w_h$, the thickness $b_h$, the attached length $l_h$, the detached length $L_h$, the elastic modulus $E_h$, the axial load within the tape attached length $P_h$, and the force transferred to the sub-level contacts $F_{h-1}$. We assume that the contact is split at the lower level (*h-1*) along the attached length of the tape in $N_h$ "rows" and $N_h$ "columns" (along *x* and *y*) of sub-level contacts (*Figure.2.A*).

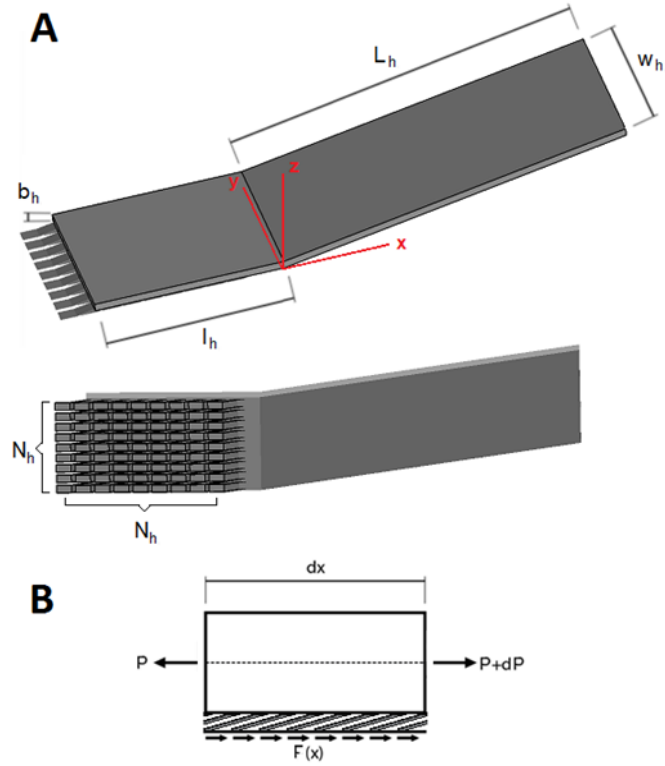

*Figure 2 : A. Schematic of the hierarchical attachment system B. Force equilibrium between two hierarchical levels.*

To simplify the analytical model, we choose a number of geometrical rules to define our hierarchical systems. First, we impose that the addition of a scale level does not reduce the total contact area, so that $l_h = N_h l_{h-1}$ and $w_h = N_h w_{h-1}$. Additionally, we apply a general "self-similar" scheme whereby all dimensions scale by the same factor between hierarchical scales, so that $b_h = N_h b_{h-1}$ and $L_h = N_h L_{h-1}$. Finally, we consider a constant elastic modulus $E$ at every scale level, which allows us to evaluate the role of pure hierarchy, although it is not necessarily realistic for some biological systems [2].

We adopt a top-down scheme to determine the load supported by each contact, starting from the larger (level-*h*) structure. The load transfer between level *h* and level (*h-1*) is obtained

from force balance on an infinitesimal length of the level $h$ attached region $dx_h$ (*Figure.2.B*), as:

$$\frac{dP_h}{dx_h} = N_h \frac{dN_h}{dx_h} F_{h-1} \tag{7}$$

where $dP_h$ is the variation of the axial load over $dx_h$ and $N_h dN_h$ is the number of contact units on the infinitesimal area $w_h dx_h$. The load transferred to level $h$-1 is assumed to be constant along the width $w_h$ of the level $h$ tape. The axial force in each contact is:

$$F_{h-1} = \frac{Eb_{h-1}w_{h-1}}{L_{h-1}} u_h \tag{8}$$

where $u_h$ is the axial displacement in the level $h$ structure. Substituting Eq. (8) into Eq. (7) and writing the strain in the level $h$ structure as $\epsilon_h = du_h/dx_h = P_h/(Eb_h w_h)$, we obtain after differentiation:

$$\frac{d^2 P_h}{dx_h^2} = \frac{P_h}{l_h L_{h-1}} \tag{9}$$

We apply the boundary condition $P_h(x_h = 0) = \widehat{P}_h$, where $\widehat{P}_h$ is the applied external load, and suppose that the length $l_h$ is sufficiently long for the axial load to tend to zero at the other tape end (as is verified in all the cases considered in this study). This is equivalent to imposing $P_h(x_h = -\infty) = 0$. We obtain from Eq. (9) the load distribution on the level $h$ as:

$$P_h(x_h) = \widehat{P}_h exp\left(\sqrt{\frac{1}{l_h L_{h-1}}} x_h\right) \tag{10}$$

From Eq. (10) we derive:

$$F_{h-1}(x_h) = \widehat{P}_h \frac{l_h}{N_h^2} \sqrt{\frac{1}{l_h L_{h-1}}} exp\left(\sqrt{\frac{1}{l_h L_{h-1}}} x_h\right) \tag{11}$$

We can then repeat the procedure iteratively for the lower levels, considering that the force applied as a boundary condition of a given contact at a given level is the force that has been transferred from the above level, i.e.:

$$\widehat{P}_h = F_h(x_{h+1}) \tag{12}$$

so that:

$$F_{h-2}(x_h, x_{h-1}) = F_{h-1}(x_h)\frac{l_{h-1}}{N_{h-1}^2}\sqrt{\frac{1}{l_{h-1}L_{h-2}}}exp\left(\sqrt{\frac{1}{l_{h-1}L_{h-2}}}x_{h-1}\right)$$

...

$$F_0(x_h, \dots, x_1) = F_1(x_h, \dots, x_2)\frac{l_1}{N_1^2}\sqrt{\frac{1}{l_1L_0}}exp\left(\sqrt{\frac{1}{l_1L_0}}x_1\right) \tag{13}$$

where $h = 0$ is the level where the tapes are in contact with the substrate, i.e. the smallest scale level.

These results are valid when the deformations within the attached regions of the level $h$ structure are small with respect to the deformation of those at level *(h-1)*. This assumption is generally valid in the study of fibrillar adhesion, since due to the elongated shape of tape-like elements and their relatively small contacts (see e.g. [35]), the displacements in the attached regions are small with respect to the ones in the detached region. If the attached length is not sufficiently long for the axial load to naturally tend to zero, Eq. (9) can be solved by imposing a boundary condition of the form $P_h(x_1 = -L_h) = 0$, which leads to an analogous exponential form for the load distribution. This case is not considered for simplicity, since we are interested in evaluating cases where maximum of detachment force is achieved, corresponding to axial loads naturally tending to zero within the contact length.

**Hierarchical load distributions**

Figure 3 shows the typical exponential contact unit load distribution for two- and three-level structures whose geometrical and mechanical properties are reported in Table 1, and applied external loads $\widehat{P_1} = 100 \ \mu N$ and $\widehat{P_2} = 3 \ mN$. In the two-level (*h=0 → h=1*) structure (Figure 3.A), e.g. the contact units adhere to the substrate and are directly attached to the tape. The exponential distribution of force transferred to the contact units presents a maximum at the peeling line ($x_1 = 0$). In the case of a three-level ($h=0 \to h=1 \to h=2$) structure (Figure.3.B), an intermediate level has been included, consisting of a set of sub-tapes. The distribution presents multiple local force maxima for each of the intermediate structures. The detachment behaviour of the first structure can easily be predicted: delamination occurs in the vicinity or the area where the load peak occurs, after which peeling proceeds at a constant pulling force (as predicted by Kendall's theory), so that a single "crack front" propagates along the substrate. All subsequent local detachment events will take place in the area adjacent to the peeling front. In the second case, the delamination events in the intermediate structures are simultaneous and several crack fronts will be involved in the detachment process. This is verified in simulations, as discussed in Section "Scaling of adhesion with hierarchical levels". In both scenarios, the force at which the system detaches is likely to be influenced by the specific overall load distribution.

| Level | $E$ | $W$ | $b$ | $L$ | $l$ | $N$ |
|---|---|---|---|---|---|---|
| 0 | 2 GPa | 200 nm | 5 nm | 0.5 μm | 200 nm | - |
| 1 | 2 GPa | 8 μm | 200 nm | 20 μm | 8 μm | 40 |
| 2 | 2 GPa | 240 μm | 6 μm | 600 μm | 240 μm | 30 |
| 3 | 2 GPa | 4.8 mm | 120 μm | - | 4.8 mm | 20 |

*Table 1: Gecko-like hierarchical structure geometrical and mechanical parameters.*

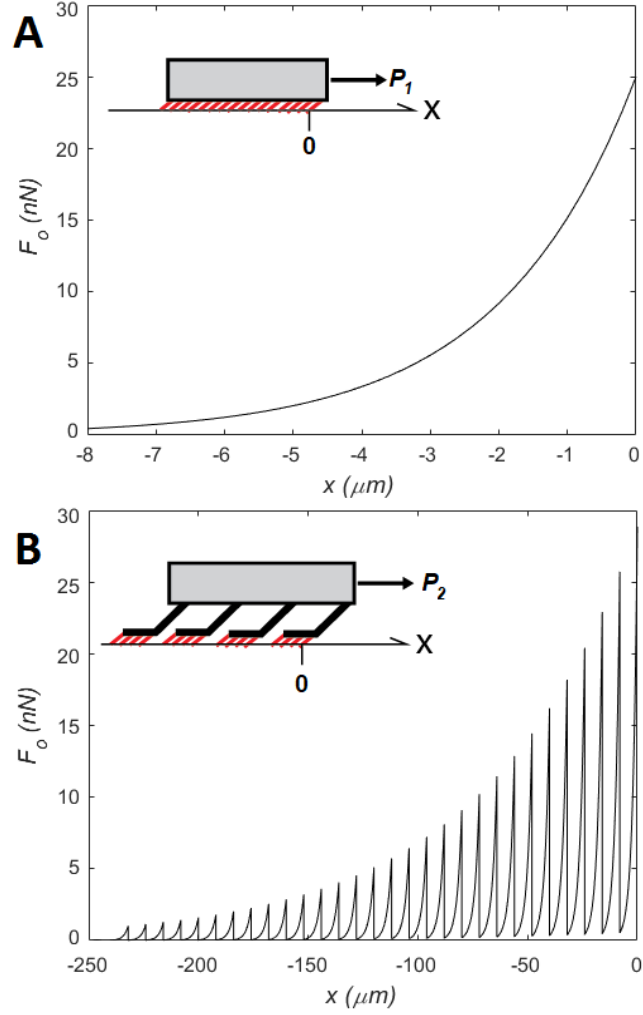

*Figure 3: Adhesion force distribution for 2-level (0-1) (A) and 3-level (0-1-2) (B) structure applying an external load $\widehat{P_1} = 100 \ \mu N$ and $\widehat{P_2} = 3 \ mN$, respectively.*

**Scaling of hierarchical adhesive energy and strength**

As discussed in Section "Thin film peeling", the energy dissipated by a detaching hierarchical structure can be obtained by considering the energy balance during delamination [28], which can be written as:

$$\frac{dW_h}{dA_h} - \frac{dU_{e,h}}{dA_h} = \frac{dU_{I,h}}{dA_h} \tag{14}$$

where $W_h$ is the work of the external force during detachment, $U_{e,h}$ is the stored elastic energy in the adhesive, $U_{I,h}$ the available energy at the interface between the adhesive and the substrate and $A_h = w_h l_h$ the attached area at level $h$. For a single-level tape, the latter is usually written in terms of critical energy release rate $G_h$ [36] as:

$$\frac{dU_{I,h}}{dA_h} = G_h \tag{15}$$

In a hierarchical adhesive structure, this can be written as the total energy that the lower scale structures can dissipate per unit area of contact before complete detachment, so that:

$$G_h = \frac{W_{h-1}}{A_{h-1}} \tag{16}$$

Thus, the total amount of dissipated energy can be obtained from Eq. (14) as:

$$W_h = \int_{A_h} \frac{W_{h-1}}{A_{h-1}} dA_h + U_{e,h} \tag{17}$$

This highlights the fact that in a hierarchical scheme, the energy balance at the upper scales depends on the total energy that the sub-scale structures can dissipate after full detachment, and not on the maximum load they can bear before detachment starts. Therefore, the stored elastic deformation at lower hierarchical levels contributes to enhanced energy dissipation. These considerations are an extension of those presented in [13], and are here applied to the detachment of a thin-film contact unit initially attached to the substrate. According to Eq. (17), the total energy dissipated by these contacts is:

$$W_0 = l_0 w_0 G_0 + (l_0 + L_0) \frac{F_{0_c}^2}{2E b_0 w_0} \tag{18}$$

Here, $G_0$ is the adhesive energy at the interface between the contact unit and the substrate, and $F_{0_c}$ is the detachment force of the contact units, which can be obtained from Kendall's equation (Eq. (2)). At the upper hierarchical scale, the available energy at the interface $G_1$ is the total amount of energy that the contacts can dissipate per unit of area (from Eq. (18)):

$$G_1 = \frac{W_0}{l_0 w_0} = G_0 + \left(1 + \frac{L_0}{l_0}\right) \frac{F_{0_c}^2}{2E b_0 w_0^2} \tag{19}$$

We can then repeat the procedure iteratively for an increasing number of levels to obtain for each the available interface energy (and therefore the detachment force, applying Kendall's energy balance):

$$G_{h+1} = \frac{W_h}{b_h w_h} = G_h + (1 + \frac{L_h}{l_h})\frac{F_{h_c}^2}{2Eb_h w_h^2} \tag{20}$$

For $\theta = 0$, i.e. the previously considered particular case of hierarchical shear lag, Kendall's equation becomes:

$$F_{h_c} = \sqrt{2Eb_h w_h^2 G_h} \tag{21}$$

Injecting Eq. (21) in Eq. (20), the scaling of the dissipated energy between levels thus becomes:

$$G_{h+1} = G_h(2 + \frac{L_h}{l_h}) \tag{22}$$

so that each additional level gives an increase in the adhesive strength by a factor of $\sqrt{2 + \beta}$, where $\beta = L_h/l_h$ is the ratio between the detached and attached length of the introduced hierarchical level "tape-like" structure. Thus, contrary to the analysis in [13], consideration of peeling in hierarchical structures leads to a scaling dependence on the ratio of the attached/detached lengths at each level. The case of nonzero peeling angles is treated below, where Eq.(21) is replaced by an angle-dependent detachment force expression

The scaling in adhesive strength in (Eq.22) corresponds to the ideal case where the introduction of a new hierarchical level does not lead to a reduction of contact area, which is not necessarily realistic, since a packing density smaller than 1 usually occurs in fibrillar interfaces. Therefore, we introduce a "packing density factor" α as the fraction of contact area Ac with respect to the available area of contact as :

$$A_{c,h-1} = \alpha A_{h-1} \tag{23}$$

Where $A_{c,h-1}$ and $A_{h-1}$ are the available contact area and the total contact area at level-h, respectively. The total energy that the lower scale can dissipate (Eq.16) thus becomes:

$$G_h = \frac{dU_{I,h}}{dA_{c,h}} = \alpha \frac{dU_{I,h}}{dA_h} = \alpha \frac{W_{h-1}}{A_{h-1}} \quad (24)$$

Using the same procedure as in Eqs. (17) to (22) leads to the following relationship between two adjacent levels:

$$G_{h+1} = \alpha G_h (2 + \frac{L_h}{l_h}) \quad (25)$$

Thus, adding realistic packing density leads to the possibility of observing a decrease in the adhesive strength as a new scale level is introduced, occurring when $\alpha < l_h/(2l_h + L_h)$. Here again, the ratio between detached and attached length of the sub-contacts is fundamental in adhesive strength optimisation.

**Numerical model**

To verify the mechanisms outlined in the previous Section, we develop a numerical procedure to simulate the complete detachment of hierarchical structures. The approach is similar to that adopted in the literature in models used to describe static and dynamic friction [37-39], although here we do not consider these aspects for simplicity. The system is discretized and modelled using a linear system of equations based on the Finite Element Method (FEM) in one dimension [40]. In particular, for a two-level system, the length $l_1$ is discretized in $n_1$ segments of length $l_1/(n_1 - 1)$ each containing $N_1^2/(n_1 - 1)$ contacts, and we add one detached segment of length $L_1$. The linear system of load-displacement equations of size $n_1^2$ is written as $= \mathbf{K} \mathbf{u_1}$, where $\mathbf{K}$ is the stiffness matrix derived using Eq. (8) and explicitly provided in the Appendix. The external load $P_1$ is applied on the terminal element of the discretized tape, so that the external force vector is $\mathbf{Q}(j) = P_1$ for $j = n_1$ and $\mathbf{Q}(j) = 0$ for $j \neq n_1$. The equilibrium is written as $\mathbf{u_1} = \mathbf{K}^{-1}\mathbf{Q}$ and the load distribution acting on each contact unit is then computed from the corresponding displacement field. For a three-level structure, the above systems are assembled over the length $l_2$ which is discretized in $n_2$ segments of length $l_2/n_2$ each of which contains $N_2^2/n_2$ sub-units, resulting in a linear system of size $(n_1 n_2)^2$. The number of levels can be increased following the same iterative procedure. The explicit form of the stiffness matrix in this case is also provided in the Appendix and the schematic of the element connectivity is shown in *Figure A.1*.

Simulations are performed by imposing a stepwise incremental displacement. An elasto-plastic force to separation law is introduced at the contact level to simulate the load response of the single contacts as well as the detachment behaviour, i.e. the initial response of these bonds is linear elastic until it reaches the theoretical peeling force from Eq. (21) and becomes perfectly plastic until full detachment occurs.

Statistical distributions are also introduced in the numerical model for the adhesive energy $G_0$ to capture the influence of surface roughness, defects and inhomogeneities, as occurs in real systems [41]. Therefore, surface energies $G_0(x_h)$ are randomly assigned for each segment along $x_h$ extracting the values from a Weibull distribution [33, 42], as shown in the inset of *Figure 4* considering various shape parameters $m$.

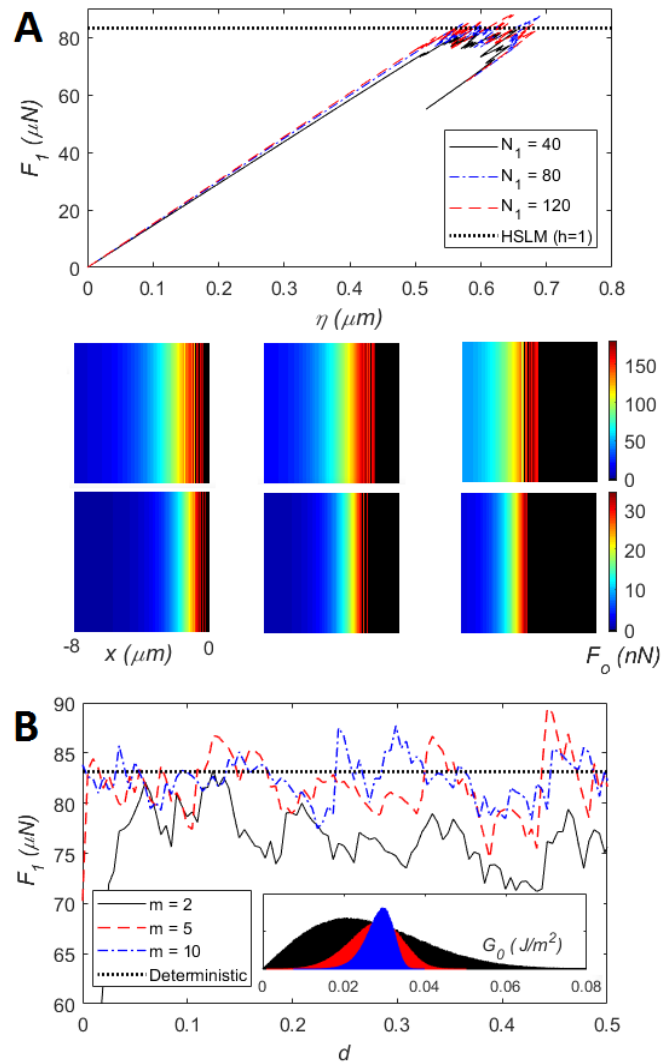

*Figure 4: A. Force vs. displacement plots during detachment for different contact array numbers and sizes. B. Maps illustrating the propagation of the peeling front during delamination for $N_1 = 40$ (first row) and $N_1 = 120$ (second row) at successive time instants t1, t2, t3. The colour scale represents the contact unit force intensity. The area where contact units are detached is displayed in black. C.*

*detachment force as a function of the ratio d between the number of fully detached contacts and the initial number of contacts for various shape parameter values m of the Weibull distribution (shown in the inset).*

**Results and discussion**

**Scaling of adhesion with contact number and size**

In order to first verify the role of fibrillar contact number and size in adhesion, simulations are performed with varying lengths and numbers of contact units. We consider a level-1 (non-hierarchical) structure, with fixed geometry and mechanical properties, and a level-2 structure with the same mechanical properties, both initially in contact with the substrate. The reference structure has the properties reported in *Table 1* (level-1), which are representative of the gecko spatula [5, 35, 43]. To evaluate the influence of the contact unit size, different values of $N_1$ are considered ($N_1=40$, $N_1=80$, $N_1=120$), allowing an increase in the total number of contacts $N_1^2$, and a reduction in their dimensions at level-0, since the total contact area is constant. An value of $G = 30$ mJ/m$^2$ is chosen, which corresponds to the typical adhesive energy between glass and a hard polymer [44]. As a first approximation, the average adhesive energy increase with the reduction of the contact tip size predicted by contact models [10] is neglected. From Eq. (21) and Eq. (22), we obtain the theoretical force at which detachment initiates as:

$$F_{1_c} = \sqrt{2Eb_1w_1^2G_0(2 + \frac{L_0}{l_0})} \qquad (26)$$

This force value is taken as the scale parameter of the Weibull distribution (*Figure 4.C*) in simulations. The numerically calculated external force $F_1$ vs. displacement $\eta$ at the load application point is shown in *Figure.4.A* for different $N_1$ values. In all cases, there is an initial linear elastic deformation phase, then the load reaches a plateau corresponding to the detachment phase.

Despite statistical variation in the local detachment forces, the average global adhesive force during detachment is relatively constant, and coincides with the theoretical value in Eq. (26). Thus, despite the increase in the total peeling line due to contact splitting, usually indicated by adhesive theories as one of the main parameters governing adhesion [1, 11], the overall detachment force is found to be constant with the number of contacts. This is due to the fact that the variation in the load distribution shown in *Figure.4.B* counteracts the effect of contact splitting, i.e. the load is distributed over a smaller fraction of the available contacts as their size decreases, so that there is no dependence of the overall detachment force with $N_1$. Only a uniformly distributed load applied to all contact units would provide an improvement in the delamination load with contact size reduction ($F_{0_C} \propto \sqrt{N_1}$). In other words, only in the ideal case of infinitely compliant contacts would contact splitting be beneficial.

*Figure.4.C* shows that the dependence of the detachment force on the chosen type of Weibull distribution is limited: for all three chosen shape parameters (governing the dispersion of the distribution) the force remains fairly constant as delamination proceeds, i.e. as function of the ratio *d* between the number of fully detached contacts and the initial number of contacts.

**Scaling of adhesion with hierarchical levels**

We now consider the level-2, level-1 and level-0 structures with the parameters given in *Table 1*, as in the case discussed in Section "Hierarchical load distributions". The adhesive energy is assigned as in the previous simulation. We also introduce a distribution for the contact unit stiffness $K_0 = E w_0 b_0 / L_0$, so that $K_0(x_h)$ are randomly assigned along the attached lenght of the adhesive system, extracting again the value form a Weibull distribution. The load response during delamination of the resulting hierarchical system is shown in *Figure 5.A* (the Weibull distribution is shown in the inset). Comparing this structure with the one obtained from the same number and dimensions of contacts, but without the intermediate level (level-1), where the analytical detachment force as in the previous simulation, an increase in the total detachment force can be observed for the 3-level structure, together with an increase in the total dissipated energy (the integral of the force vs. displacement curve). Due to the particular shape of the load distribution within the hierarchical system, more contacts are involved during in the detachment process, resulting in an increased overall detachment force. As previously, an analytical force at which detachment occurs can be calculated from Eq. (21) and Eq. (22) as follows:

$$F_{2_C} = \sqrt{2 E b_2 w_2^2 G_0 \left(2 + \frac{L_1}{l_1}\right)\left(2 + \frac{L_0}{l_0}\right)} \qquad (27)$$

This load level is also plotted in.        x (μm)

*Figure 5.A,* showing good agreement with numerical simulations. The increase in adhesive strength can be explained by the fact that the detachment process involves the creation of multiple "crack fronts", as illustrated in *Figure 5.B*, which is beneficial to the overall adhesive performance. We note that the hierarchical load distribution is observed even with the random distributions introduced on detachment energies and contact unit stiffnesses. As shown in Figure 5, the load distribution during the 3-level peeling test displays some noise due to the statistical distributions, but can still be fitted by the theoretical distribution obtained in the deterministic case. Results in terms of global detachment force remain relatively insensitive to the local variations at the contact level, as in the previous simulations. However, the present model only considers one- dimensional effects. In a more realistic scenario, two- and three-dimensional load concentrations as a result of imperfect contact could lead to a decrease in the overall detachment force. As the system starts to detach, an equilibrium between the propagation of different crack fronts is reached. These results confirm that the maximum load that an adhesive structure can bear is related principally to the energy that can be dissipated by its interfacial contacts rather than to their delamination strength. In other words, the increase in detachment strength is mainly due to the increase in adhesive energy occurring at each additional hierarchical level. Additionally, these results highlight the fact that as the contact sizes become critical, biological adhesives adopt hierarchical organization to maintain the presence of multiple peeling fronts over the whole length of the attached system, giving rise to optimized distributions and developing a maximal delamination force from a given overall contact area.

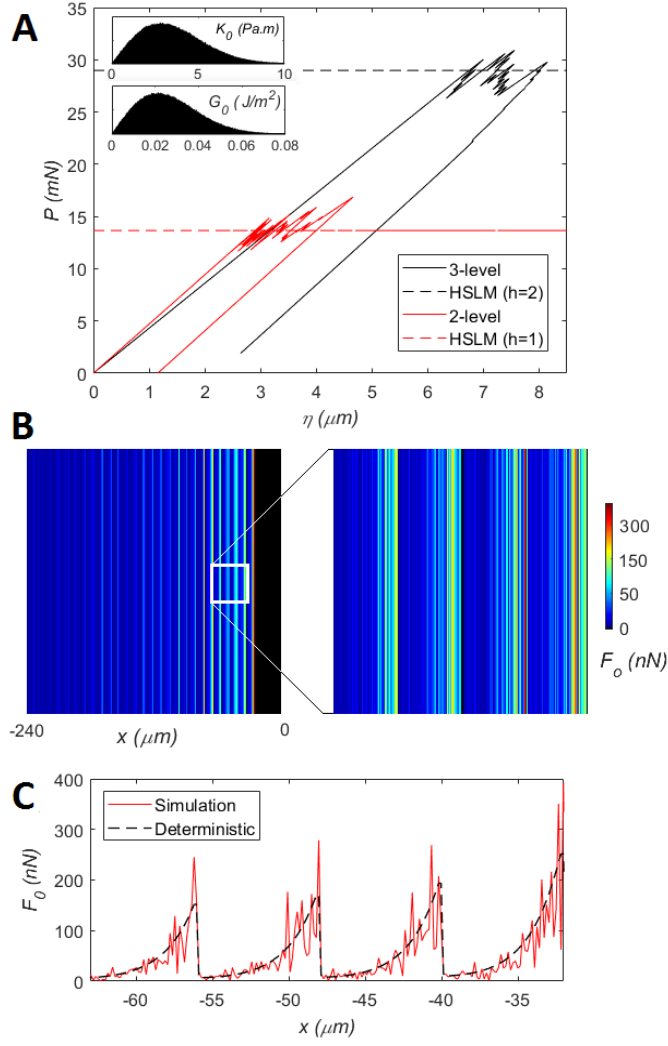

*Figure 5 : A. Force vs. displacement curves for 2-level and 3-level structures. B. Propagation of multiple peeling fronts during simulation of the 3-level structures. C. Adhesion force distribution including statistical distributions over the contact length of the system.*

The application of the proposed model for contact load distribution within a hierarchical attachment geometry is limited to the transfer between axial load in the tape and tangential forces within the interface, but the proposed method to extract the scaling of adhesive strength from the energy balance can be extended to different loading cases. Let us consider for example peeling under pure bending as described in [45], where a rotation is applied ad the end of the detached length rather than a force parallel to the substrate. In this case, the elastic energy term in Eq.(17) no longer corresponds to the tensile strain energy due to axial forces, but is due to the bending of the tape, so that:

$$W_h = l_h w_h G_h + (l_h + L_h)\frac{6M_{h_c}^2}{E b_h^3 w_h} \qquad (28)$$

where $M_{hc}$ is the applied bending moment when detachment occurs. At the upper level, the available energy at the interface becomes:

$$G_{h+1} = G_h + (1 + \frac{L_h}{l_h})\frac{6M_{h_c}^2}{E b_h^3 w_h^2} \qquad (29)$$

Introducing the relationship between the critical bending moment and the geometrical and mechanical properties of the system [45]:

$$M_{h_c} = \sqrt{\frac{E b_h w_h^2 G_h}{6}} \qquad (30)$$

The scaling of adhesive strength when a new hierarchical level is introduced becomes:

$$G_{h+1} = G_h(1 + \frac{1}{b_h^2} + \frac{L_h}{b_h^2 l_h}) \qquad (31)$$

Here, in contrast to the shear lag case, the tape thickness strongly influences the adhesive properties of the hierarchical system. In a more general case, the peeling of the adhesive tape involves tensile, shear and bending strain energies, which are stored in the interface as the detachment propagates. Since the analytical calculation of the critical tensile, shear and bending load and the corresponding strain energies stored in the sub contacts as a function of the tape geometries are difficult to find in closed form (see [28]), numerical models as the one presented here are useful for the calculation and optimization of hierarchical systems when mixed loading conditions are considered.

**Peeling angle-dependency of hierarchical tape arrangements**
In terms of load distributions, both normal and tangential loads are present for peeling angles greater than zero. For the tangential component, the distribution remains the same as for the zero-angle case discussed in Section "Hierarchical Shear-Lag Model", with reduced amplitudes. For the normal component, a closed-form analytical solution cannot be derived, since it would require the solution of a nonlinear system of equations. However, Eq. (20) can be generalized using Kendall's theory (Eq. (2)) to the detachment of a thin film at a peeling angle $\theta$, and the relationship between the detachment force and available interface energy at a given level $h$ can be written as:

$$F_{h_c} = Eb_h w_h \left( \cos\theta - 1 + \sqrt{(1 - \cos\theta)^2 + \frac{2G_h}{Eb_h}} \right) \tag{32}$$

This expression replaces Eq. (21) in the case of nonzero angle peeling.

Starting from level-0, the detachment force of each contact unit is calculated as a function of the contact interface energy. The upper level available energy and detachment forces are then iteratively calculated following this scheme in order to derive the overall detachment force. We apply this iterative procedure to the whole structure from *Table 1* (level-0, level-1, level-2 and level-3). *Figure.6.A* illustrates the scaling of available adhesive energy at the interface $G_h$ for each considered level as a function of the peeling angle. A clear advantage of a hierarchical arrangement with multiple levels is highlighted in terms of energy dissipated by the "hierarchical interface" at small angles.

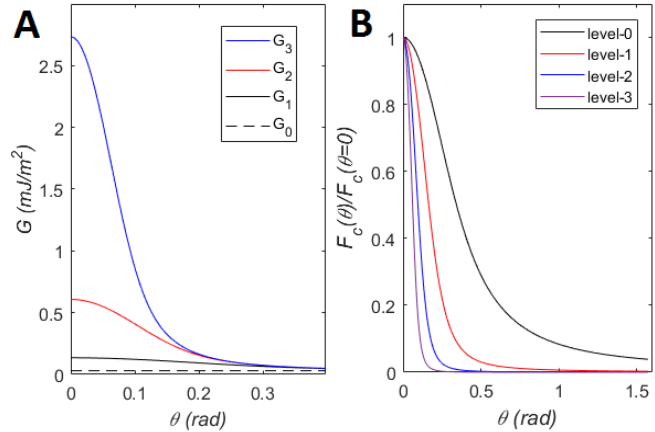

*Figure 6 A. Scaling of the adhesive energy of hierarchical self-similar tape structures: A) Overall adhesive strength as a function of peeling angle for 2-level (1.000.000:1), 3-level (1.000:1.000:1) and 4-level (100:100:100:1) structures with constant overall number of contacts. B. Overall adhesion force vs. peeling angle $\theta$ for the three structures, normalized with respect to the $\theta = 0$ value.*

As the peeling angle increases, the available energy at each level tends to that at the contact level $G_0$, so that no improvement is obtained from structural features. The angle dependency is that found in single-peeling theory, and results shows that the efficiency of the hierarchical structures is also angle-dependent, as shown in *Figure.6.B*.

## Conclusions

In conclusion, we have developed a generalization of the shear lag model to describe hierarchical fibrillar systems such as those observed in gecko and arachnid attachments and applied it in numerical simulations. We have shown that improved adhesion in fibrillar structures is not simply due to contact splitting alone, but rather to hierarchical organization, giving rise to optimized load distributions, enabling reduced stress concentrations, and therefore a reduced risk of detachment. In fact, we show that the effect of contact splitting, which was originally derived for punch-like geometries using a JKR model [10] and discussed in detail in [12], is counterbalanced by the effect of load concentrations in the case of tangential tape peeling, and therefore is not beneficial for increasing adhesion in the absence of hierarchical structure, or in an ideal case of extremely compliant contacts. These results are consistent with those obtained with other approaches such as the spring-block model in the case of static friction [46]. Hierarchical architectures are shown to provide the means to generate multiple delamination fronts once detachment initiates, and therefore to increase energy dissipation and adhesive strength. The general scaling behaviour of the adhesion of hierarchical structures is discussed for constant and reduced contact areas, showing a clear advantage in providing multiple hierarchical levels. These mechanisms could help explain results such as those reported in [47], where an increase in animal adhesive pads' adhesive efficiency with size, for which the mechanism is still unclear, is observed. Both the calculated pull-off forces (in the 50 $\mu$N to 50 mN range for an increasing number of hierarchical levels) and the gain in adhesive strength at each hierarchical level (from 50% to 150%), obtained for typical geometrical parameters such as those in Table 1, are compatible with existing numerical [12] and experimental [48] results on hierarchical adhesives. The presented model and numerical analysis provide for the first time an evaluation of the influence of load distributions and simultaneous delamination fronts in peeling problems, and the study contributes to providing a better understanding of the mechanisms of adhesion of hierarchical structure. Results can be used to provide design and optimization criteria for artificial adhesive structures, and possibly for optimized composite materials with hierarchical reinforcements [49].

## Conflicts of interest

There are no conflicts to declare.

## Acknowledgements

N.M.P. is supported by the European Commission H2020 under the Graphene Flagship Core 1 No. 696656 (WP14 "Polymer composites") and FET Proactive "Neurofibres" grant No. 732344. F.B. is supported by H2020 FET Proactive "Neurofibres" grant No. 732344 and by project "Metapp", (n. CSTO160004) cofounded by Fondazione San Paolo. This work was carried out within the COST Action CA15216 "European Network of Bioadhesion Expertise: Fundamental Knowledge to Inspire Advanced Bonding Technologies". Computational resources were provided by the C3S centre at the University of Torino (c3s.unito.it) and hpc@polito (www.hpc.polito.it).

## Appendix

### Equations for the numerical model

For a two-level structure, the linear system of equations for the FEM simulations is banded and of size $n_1^2$:

$$\mathbf{K} = \begin{bmatrix} k_0 + k_1 & -k_1 & 0 & \cdots & & \cdots & 0 \\ -k_1 & k_0 + k_1 & \ddots & \ddots & & \cdots & \vdots \\ 0 & \ddots & \ddots & \ddots & & 0 & \vdots \\ \vdots & & \ddots & \ddots & k_0 + k_1 & -k_1 & 0 \\ \vdots & \cdots & & 0 & -k_1 & k_0 + k_1 + k_{1d} & -k_{1d} \\ 0 & \cdots & & & 0 & -k_{1d} & k_{1d} \end{bmatrix}$$

(A.1)

where $k_1 = n_1 E_1 b_1 w_1 / l_1$, $k_{1d} = E_1 b_1 w_1 / L_1$ and $k_0 = (N_1^2 E_0 b_0 w_0)/(n_1 L_0)$.

For a three-level structure, we first build the stiffness matrix corresponding to the contribution in the linear system of the level-0 and level-1:

$$\mathbf{K_1} = \begin{bmatrix} \mathbf{K} & 0 & \cdots & 0 \\ 0 & \mathbf{K} & \ddots & \vdots \\ \vdots & \ddots & \ddots & 0 \\ 0 & \cdots & 0 & \mathbf{K} \end{bmatrix}$$

(A.2)

The sub-matrixes in the above matrix are obtained from A.1 with $k_1 = (n_1 N_2^2 E_1 b_1 w_1)/(n_2 l_1)$, $k_{1d} = (N_2^2 E_1 b_1 w_1)/(n_2 L_1)$ and $k_0 = (N_1^2 N_2^2 E_0 b_0 w_0)/(n_1 n_2 L_0)$.

We then add the Level-2 contribution:

$$K_{2\,ij} = \begin{cases} k_2 & \text{for } (i = j = 1) \cup (i = j = n_1 n_2) \\ 2k_2 & \text{for } (i = j = p n_1) \cap (i \neq n_1) \cap (i \neq n_1 n_2) \quad p \in \mathbb{N} \\ -k_2 & \text{for } (i = p n_1) \cap (i = j \pm p n_1) \end{cases}$$

(A.3)

with $k_2 = n_2 E_2 b_2 w_2 / L_2$.

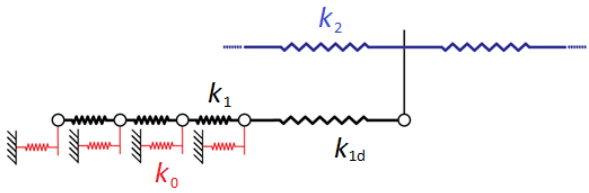

*Figure A.1 Schematization of the hierarchical connectivity of elements (corresponding to the adopted stiffness matrix) used in the simulations.*